\documentclass[twocolumn,showpacs,preprintnumbers,amsmath,amssymb]{revtex4}
\usepackage{graphicx}% Include figure files
\usepackage{dcolumn}% Align table columns on decimal point
\usepackage{bm}% bold math
\usepackage{color}
\usepackage{ulem}
\begin{document}
\preprint{APS/123-QED}

\title{Transmission channels for light in absorbing random media: from diffusive to ballistic-like transport}% Force line breaks with \\

\author{Seng Fatt Liew}
\affiliation{Applied Physics Department, Yale University, New Haven CT 06520, USA.}
\author{S\'{e}bastien M. Popoff}
\affiliation{Applied Physics Department, Yale University, New Haven CT 06520, USA.}
\author{Hui Cao}
\email{hui.cao@yale.edu}
\affiliation{Applied Physics Department, Yale University, New Haven CT 06520, USA.}
\author{Allard P. Mosk}
\affiliation{Complex Photonic Systems (COPS), $MESA^+$ Institute for Nanotechnology, University of Twente, P.O. Box 217, 7500 AE Enschede, The Netherlands.}
\author{Willem L. Vos}
\affiliation{Complex Photonic Systems (COPS), $MESA^+$ Institute for Nanotechnology, University of Twente, P.O. Box 217, 7500 AE Enschede, The Netherlands.}%

\date{\today}% It is always \today, today,

\begin{abstract}
While the absorption of light is ubiquitous in nature and in applications, the question remains how absorption modifies the transmission channels in random media. 
We present a numerical study on the effects of optical absorption on the maximal transmission and minimal reflection channels in a two-dimensional disordered waveguide. 
In the weak absorption regime, where the system length is less than the diffusive absorption length, the maximal transmission channel is dominated by diffusive transport and it is equivalent to the minimal reflection channel. 
Its frequency bandwidth is determined by the underlying quasimode width. 
However, when the absorption is strong, light transport in the maximal transmission channel undergoes a sharp transition and becomes ballistic-like transport. 
Its frequency bandwidth increases with absorption, and the exact scaling varies with the sample's realization. 
The minimal reflection channel becomes different from the maximal transmission channel and becomes dominated by absorption. 
Counterintuitively, we observe in some samples that the minimum reflection eigenvalue increases with absorption. 
Our results show that strong absorption turns open channels in random media from diffusive to ballistic-like.
\end{abstract}

 %Electronic transport in mesoscopic systems%% PACS, the Physics and Astronomy
\maketitle

\section{\label{sec:level1}Introduction}

\indent In mesoscopic transport, wave interference plays an essential role, giving rise to well-known phenomena such as enhanced backscattering, Anderson localization and universal conductance fluctuations \cite{van_albada,wolf,altshuler1991mesoscopic,maret,Akkermans_book, Lagendijk_PhysToday}. 
Recently, another striking interference effect has caught much attention, that is, the existence of highly transmitting channels, terms ``open channel'' in a random system \cite{vellekoop_PRL, Pendry_view, choi_PRB, choi_OE, shi_PRL12, choi_NP, Davy_OE13, sebastien13}. 
These open channels, which enable an optimally prepared coherent input beam to transmit through a strong scattering medium with order unity efficiency, were predicted initially for electrons \cite{dorokhov1,dorokhov2,Mello88, Nazarov94}. 
However, they have not been directly observed in mesoscopic electronics due to the extreme difficulty of controlling the input electron states. 
In contrast, it is much easier to prepare the input states of classical waves, such as electromagnetic waves or acoustic waves. 
Recent developments of adaptive wavefront shaping and phase recording techniques in optics have enabled experimental studies of open channels \cite{vellekoop_OL, changhueiyang_NP, sebastien_PRL, mosk_NP}. 
The open channels greatly enhance light penetration into scattering media, that will have a profound impact in a wide range of applications from biomedical imaging and laser surgery to photovoltaics and energy-efficient ambient lighting \cite{sebastien_NC, mosk_NP, vos2013}. \\

\indent The transmission channels are eigenvectors of the matrix $t^{\dagger}t$, where $t$ is the field transmission matrix of the system. 
The eigenvalues $\tau$ are the transmittance of the corresponding eigenchannels. 
In the lossless diffusion regime, the density of eigenvalues $\tau$ has a bimodal distribution, with one peak at $\tau \simeq 0$ that corresponds to closed channels, and a peak at $\tau \simeq 1$ that corresponds to open channels \cite{dorokhov1,dorokhov2,Mello88, Nazarov94}. 
The diffusion process is dominated by the open channels, and the average transmittance is proportional to the ratio of the number of open channels by the total number of propagating channels \cite{dorokhov1,dorokhov2}. 
At the transition to localization, the number of open channels is reduced to one, and the open channels disappear in the localization regime. 
The conductance of a localized system is dominated by the single highest transmitting channel. 
In the past few years, wavefront shaping has been utilized to increase the coupling of the incident light to the open channels of random media \cite{vellekoop_PRL, choi_NP, shi_PRL12, kim_OL13, sebastien13, choi13a}. 
Numerical simulations reveal that the open channels enhance the energy stored inside the disordered medium \cite{choi_PRB}. 
In the diffusion regime, the field energy of an open channel is spread over the entire transverse extent of a sample, while in the localization regime the maximal transmission channel becomes confined in the transverse direction normal to the transmission direction \cite{choi_OE}. \\

\indent In addition to the transmission channels, the transport can also be interpreted in terms of resonances, which are referred to as ``levels'' for electrons and ``modes'' for classical waves \cite{genack_nature, genack_JMP}. 
For an open system, one can define quasi-normal modes of the Maxwell equations. 
They describe states that have stationary normalized spatial profiles and amplitudes decaying in time due to radiative losses. 
These quasi-normal modes play an important role in transport, e.g., in the localization regime energy is transported either by tunneling through a localized mode in the middle of the sample or by hopping over a necklace state that is formed via coupling of several localized modes \cite{pendry, pendry_AP94, bertolotti_PRL05, sebbah_PRL, bliokh_PRL, bliokh_PRL08}. 
If the input beam is coupled to multiple modes, the interference of these modes at the output end determines the total transmission \cite{genack_nature}. 
The two approaches to describe transport phenomena, transmission channels and resonant modes, complement each other \cite{genack_JMP}. 
At any frequency, the transmission eigenchannels can be expressed as a frequency-dependent superposition of resonant modes with specific resonance frequencies and widths. 
In the localization regime, the maximal transmission channel can typically be identified with a single resonant mode \cite{genack_JMP}. \\

\indent Absorption of radiation is usually assumed to suppress interference effects such as the occurrence of open channels. 
Most studies on transmission eigenchannels have considered lossless random media where absorption is negligible. 
In reality absorption exists in any material system, and could have a significant impact on diffusion and localization \cite{brouwer, yamilov_OE13}. 
In the microwave regime for instance, absorption is particularly difficult to avoid and leads to a significant reduction of the transmission through disordered waveguides \cite{shi_PRL12,chabanov_Nat00}. 
Absorption does not destroy the phase coherence of scattered waves, but it attenuates the longer scattering paths more than the shorter paths, thus modifying the interference patterns. 
Since the open channels penetrate deeper into the random medium than the closed channels, they would experience more absorption. 
In other words, absorption should have a stronger effect on the open channels than on the closed channels. 
However, it is not clear how absorption would modify the open channels. 
Moreover, it has been shown lately that light absorption in strongly scattering media can be greatly enhanced or inhibited by coherent effects \cite{chong_PRL10, wan_Sci11, chong_PRL11}. 
Thus the interplay between absorption and interference determines not only the amount of energy being transmitted, but also the amount of energy being deposited in the random media. 
The investigation of strongly scattering systems with absorption is therefore not only important to the fundamental study of mesoscopic transport, but also relevant for applications in imaging, light harvesting, and lighting technology \cite{sebastien_NC, mosk_NP, vos2013,gratzel_JPPC03,hagfeldt_CL10}.\\

\indent In this paper, we address the following questions: 
how does absorption modify the open channels? 
How does the channel bandwidth vary with absorption? 
What is the correlation between a transmission and a reflection channel in the presence of absorption? 
In weak absorption, when the ballistic absorption length $l_a$ is much larger than the average path length $l_p = 2L^2/l_t$ ($l_t$ is the  transport mean free path), most scattering paths are not affected by absorption. 
However, when ballistic absorption length becomes smaller than the average path length $l_a < l_p$, attenuation of long scattering paths significantly affects the transport through the system. 
To study the change of light transport, we compute the spatial field distribution inside the random medium. 
The spectral width of the maximal transmission channel is important to many of the aforementioned applications, for instance, a broad spectral width is desired for light harvesting. 
To address this question, we calculate the frequency bandwidth of input wavefront corresponding to the maximal transmission channel and its scaling with absorption. 
Experimental studies of the transmission channels rely on the access to both sides of the scattering media. 
It is, however, often more convenient and less invasive to work in a reflection configuration, where all measurements are on the input side of the sample. 
For example, without absorption, the total transmission is maximized by finding the minimal reflection channel \cite{choi2013b}. 
In presence of absorption, the relation between the maximal transmission channels and the minimal reflection channels is not known. 
Therefore, we investigate the correlation between these two channels as a function of absorption. \\

\indent This paper is organized as follows. 
In section II, we present our numerical model of two-dimensional disordered waveguides. 
In section III we show how absorption modifies the maximal transmission channel and we illustrate its correlation with the quasi-normal modes. 
In Section IV, we investigate the minimal reflection channel in the absorbing random media. Section V is the conclusion.

\section{\label{sec:level2} Numerical model}
\begin{figure}[htbp]
\centering
\includegraphics[scale=0.18]{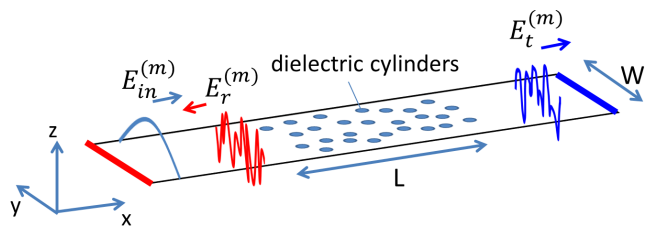}% Here is how to import EPS art
\caption{(Color online) Schematic of the 2D disordered waveguide used in our numerical simulation. Dielectric cylinders are placed randomly in a waveguide with perfect-reflecting sidewalls. The waveguide width is $W$, and the length of the disordered region is $L$. A light field $E_{in}^{(m)}$ is launched from the left end of the waveguide, and scattered by the cylinders. The transmitted light field $E_t^{(m)}$ is probed at the right end, and the reflected light field $E_r^{(m)}$ at the left end. Perfectly-matched-layers are placed at both open ends to absorb the transmitted and reflected waves.}
\label{Figure_1}
\end{figure}

We consider a two-dimensional (2D) disordered waveguide, shown schematically in Fig. \ref{Figure_1}. 
Dielectric cylinders with refractive index $n=2.5$ and radius $r_c=0.098\lambda$ are randomly positioned inside the waveguide with perfectly reflecting sidewalls. 
The dielectric cylinders occupy an area fraction of 0.04 corresponding to an average distance between cylinders of $a=0.87 \lambda$. 
We select to work at the wavelength of input light that avoids the Mie resonances of individual dielectric cylinders. 
This frequency is in photonic regime above the first band gap of a triangular lattice with the same area fraction. 
Light enters the waveguide from the left open end and is scattered by the cylinders. 
Light transmitted through or reflected from the random array is absorbed by the perfectly-matched-layers placed at both ends of the waveguide. 
We consider transverse-magnetic (TM) polarized light, whose electric field is parallel to the cylinder axis ($z$-axis). 
The width of the waveguide is $W = 10.3\lambda$, the number of guided modes in the empty waveguide is $N = 2W/\lambda  = 20$. 
The length of the random array of cylinders is $L = 20.2\lambda$.\\

\indent To calculate the electromagnetic field inside the random waveguide, we solve  Maxwell's equations using the finite-difference frequency-domain method \cite{comsol_42a}. 
The intensity is averaged over a cross-section of the waveguide to obtain the evolution $I(x)$ along the waveguide in the $x$ direction. 
The ensemble-averaged $I(x)$ exhibits the well known linear decay, from which we extract the transport mean free path $l_t = 0.073L$ \cite{van_rossum}. 
Since $l_t \ll L$, light experiences multiple scattering. 
The  localization length is $\xi = (\pi /2) N l_t = 2.3L$. 
The system is in the diffusion regime, as confirmed from the linear decay of intensity but it is close to the localization regime. 
The reason we chose this regime is as follows. 
In the localization regime, the maximal transmission eigenchannel is composed of only one or two quasi-normal modes \cite{genack_JMP}. 
In the diffusion regime ($\xi \gg L$), many overlapping modes contribute to the maximal transmission channel, making the analysis complicated. 
In our case, the maximal transmission channel consists of a few quasi-normal modes, their interference determines the behavior of the transmission channel. 
Moreover, in this regime the transport displays a large fluctuation from one realization to another. 
%allowing us to study individual realizations that are more localized or more diffusive the typical sample in the ensemble. 
Within the same ensemble we may find realizations that are more localized or more diffusive than the typical sample. 
We can therefore study a wide range of conditions in the same ensemble. \\

\indent Usually, absorption exists either inside the scattering particles, in the waveguide wall or in the background material that hosts the particles. 
The concomitant contrast in the imaginary part of the refractive index causes additional scattering, which modifies the resonant modes \cite{wu_JOSAB07}. 
In this paper we prefer to avoid this scattering effect by introducing a spatially homogeneous imaginary refractive index $\gamma$ to both scatterers and background, so that mode wavefunctions do not change and we can focus on the effects of absorption and energy loss.
The ballistic absorption length is $l_a = 1/(2k\gamma)$, where the wavevector is $k = 2 \pi/ \lambda$. 
When the ballistic absorption length $l_a$ reaches the average path length of light in a 2D diffusive system $l_p = 2 L^2 / l_t$, the diffusive absorption length $\xi_a = \sqrt{l_tl_a/2}$ becomes equal to the system length $\xi_a = L$.   \\

\indent To construct the transmission matrix $t$ of the disordered waveguide, we use the guided modes or propagation channels in the empty waveguide as the basis. 
We launch a guided mode $E_{in}^{(m)}$ from the input end, calculate the transmitted wave and decompose it by the empty waveguide modes at the output end, $E_t^{(m)} = \sum\limits_{n=1}^{N}t_{nm} E_{in}^{(m)}$. 
The coefficient $t_{nm}$ relates the field transmission from an input channel $m$ to an output channel $n$. 
After repeating this procedure for $m = 1, 2, ... N$, we obtain all the elements $t_{nm}$ for the transmission matrix $t$.
Similarly, the reflection matrix is constructed by computing the reflected waves $E_r^{(m)}$ at the input end. \\

\section{\label{sec:level3} Maximal transmission channel}
\indent A singular value decomposition of the transmission matrix $t$ gives
\begin{eqnarray}
t = U\  \Sigma \ V^{\dagger} \ , 
\end{eqnarray} 
where $\Sigma$ is a diagonal matrix with non-negative real numbers,  $\sigma_n = \sqrt{\tau_n}$, $\tau_n$ is the eigenvalue of $t^{\dagger} t$, $\tau_1 > \tau_2 > \tau_3 ... > \tau_N$. 
$U$ and $V$ are $N \times N$ unitary matrix, $V$ maps input channels of the empty waveguide to eigenchannels of the disordered waveguide, and $U$ maps eigenchannels to output channels. 
The column vectors in $V$ ($U$) are orthonormal and are called input (output) singular vectors. 
The value $\tau_n$ represents the transmittance of the $n^{th}$ transmission channel.
The input singular vector corresponding to the highest transmission eigenvalue $\tau_1$ gives the maximal transmission eigenchannel, its elements represent the complex coefficients of the waveguide modes that combine to achieve maximum transmission through the random medium.\\

\subsection{\label{sec:sublevel1} Effects of absorption on spatial field distribution and energy flow of the maximal transmission channel}
\begin{figure*}
\centering
\includegraphics[scale=0.38]{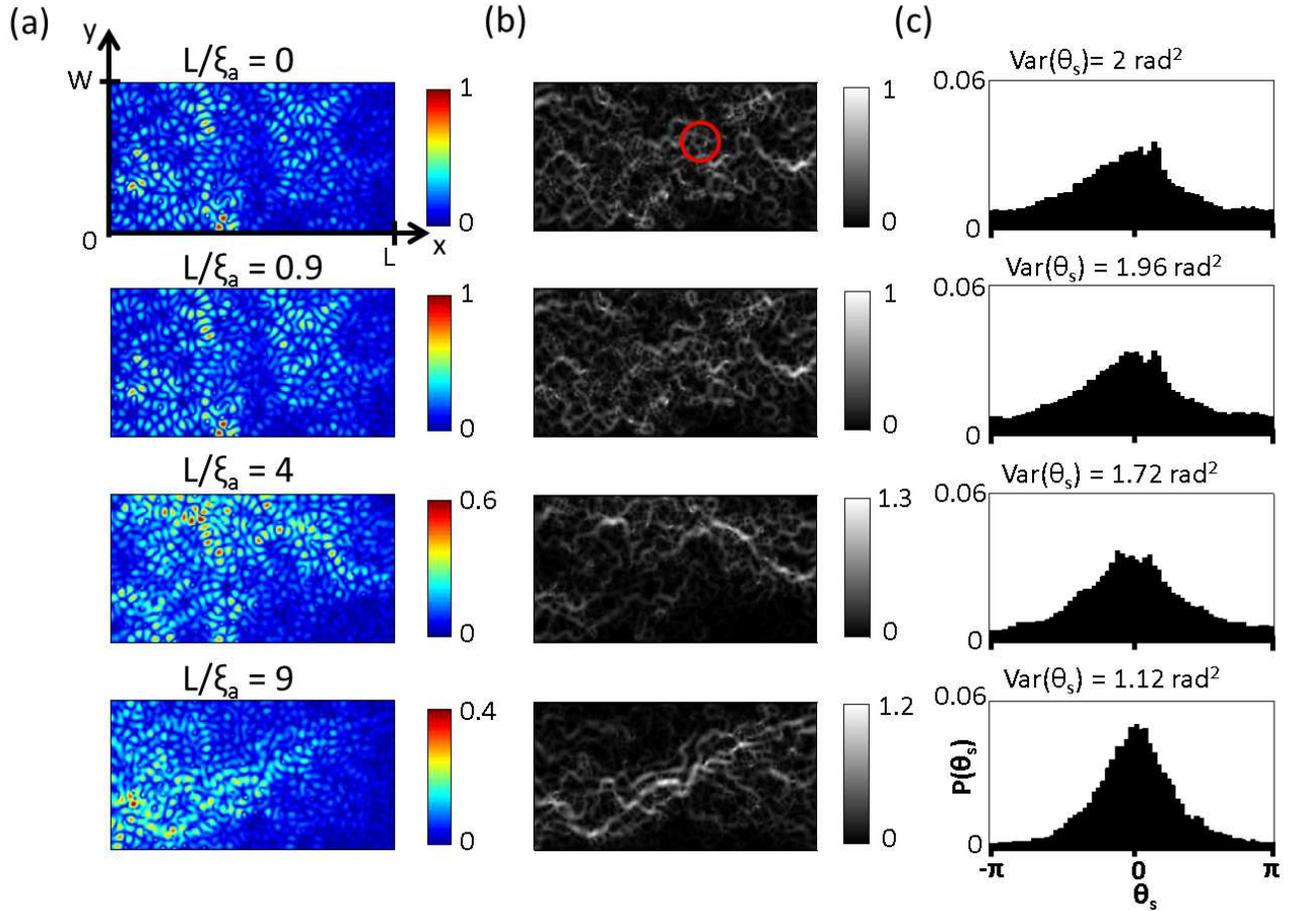}
\caption{(Color online) Evolution of maximal transmission channel with absorption. Calculated electric field amplitude $|E_z|$ [column (a)], normalized Poynting vector amplitude $|\vec{S}'(x,y)|$ in gray scale [column (b)], and histogram of weighted Poynting vector direction $P(\theta_s)$  [column (c)] of the maximal transmission channel inside the disordered waveguide as absorption $L/\xi_a$ increases from top to bottom. In (a), the maximal transmission channel remains robust against absorption with nearly identical field pattern up to $L = \xi_a$ and changes significantly beyond that point. In (b) the winding paths of light are illustrated inside the random structure in the weak absorption regime $L/\xi_a < 1$, and more straight ``snake-like'' paths in the strong absorption regime $L/\xi_a > 1$.  A "loop" in the energy flow is circled in red. In (c), the angle $\theta_s$ from the Poynting vector to the $x$-axis is widely spread between $-\pi$ and $\pi$ when $L/\xi_a < 1$, but concentrates close to 0 when $L/\xi_a > 1$. The variance of $\theta_s$, indicated above each panel, decreases with increasing $L/\xi_a$ } 
\label{Figure_2}
\end{figure*}

We inject light into the maximal transmission channel and investigate the field profile inside the random medium. 
In Fig. \ref{Figure_2}(a) we plot the spatial distribution of the electric field amplitude $|E_z|$ inside the disordered waveguide with increasing $L/\xi_a$.  
To map the energy flow inside the disordered medium, we compute the Poynting vector $\vec{S}(x,y) = \frac{1}{2}\textnormal{Re}[\vec{E}(x,y)\times \vec{H}^*(x,y)$].
Its projection onto the propagation direction ($x$-axis) is $S_x(x,y) = \vec{S}(x,y) \cdot \vec{e}_x$, where $\vec{e}_x$ is the unit vector along the $x$-axis. 
The net flow over a cross section of the disordered waveguide is $F(x) = \int_0^W S_y(x,y) dy$. 
Without absorption, the net flow is a constant, $F(x) = F(0)$.
In the presence of absorption, $F(x)$ decays exponentially along $x$. 
For a clear visualization of the energy flow deep inside the random structure, we normalize the Poynting vector $\vec{S}(x,y)$ by $F(x)$ to compensate the energy decay such that  $\vec{S}'(x,y) = \vec{S}(x,y)/F(x)$.
Figure \ref{Figure_2}(b) plots the magnitude of normalized Poynting vector $|\vec{S}'(x,y)|$. 
For a quantitative analysis of the light propagation direction, we compute the angle of the Poynting vector $\vec{S}(x,y)$ with respect to the $x$-axis, $\theta_s(x,y) = \tan^{-1}[(\vec{S}(x,y) \cdot \vec{e}_y) / (\vec{S}(x,y) \cdot \vec{e}_x)]$ where  $\vec{e}_y$ is the unit vector along the $y$-axis.  
In Fig. \ref{Figure_2}(c), we plot the histogram of $\theta_s$ weighted by the relative amplitude of the Poynting vector,
$P(\theta_s) = \int |\vec{S}'(x,y)| \delta(\theta_s - \theta_s(x,y)) dx dy$. \\
 
% % % re-organize the following paragraphs 
\indent Let us now discuss the results in Fig. \ref{Figure_2}(a-c). 
When absorption is weak ($L/\xi_a < 1$), the maximal transmission channel has nearly the same field pattern as the channel without absorption. 
The energy flow inside the random structure resembles meandering ``creeks'' that are intertwined, and many ``loops'' are seen. 
The multiply scattered light propagates in many directions, and the distribution of Poynting vector's angle $P(\theta_s)$ is broad and has a large variance.      
Once $L/\xi_a$ exceeds 1, the spatial profile evolves.
At $L/\xi_a = 4$, a noticeable change of the field pattern is observed: the loops gradually disappear, and the creeks become straighter. 
This behavior occurs because the longer scattering paths that involve more windings are strongly attenuated by absorption. 
To maximize the transmission through the random system, light takes a shorter and more straight path to minimize absorption. 
As a result, the distribution of Poynting vector's angle $P(\theta_s)$ becomes narrow and its variance decreases. 
When absorption becomes very strong ($L/\xi_a = 9$), the maximal transmission channel bears no resemblance to the one with weak absorption. 
All meandering creeks eventually merge into a single stream with few windings and light propagates mostly in the forward direction. 
We conclude that in the presence of strong absorption, diffusive transport has been turned into a ballistic-like transport, even in a thick mesoscopic sample with a length close to the localization length. 
This ballistic-like transport may enable new modes of imaging that are specific to absorbing media.\\
 
\begin{figure}
\centering
\includegraphics[scale=0.23]{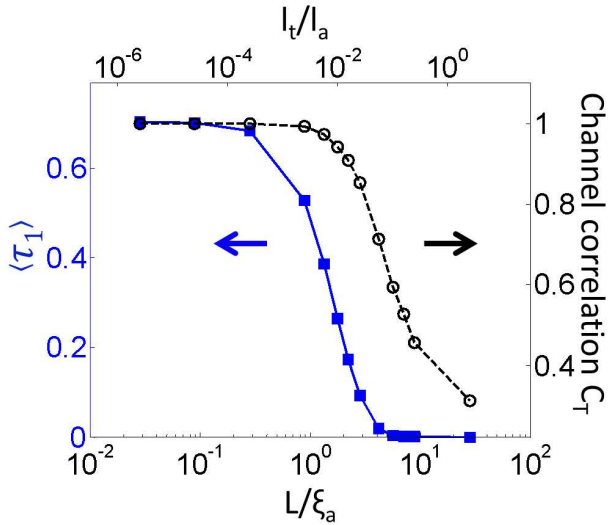}
\caption{(Color online) Maximal transmission eigenvalue and eigenvector versus absorption. Filled squares connected by solid line represent the ensemble-averaged highest transmission eigenvalue $\langle \tau_1 \rangle$ as a function of absorption (bottom axis $L/\xi_a$, top axis $l_t /l_a$). Open circles connected by dashed line represent the correlation of the input singular vector with absorption to the one without absorption $C_T$. The maximal transmission eigenchannel changes at higher absorption level compared to its eigenvalue.}
\label{Figure_3}
\end{figure}

\indent For a quantitative characterization of the change of the maximal transmission channel by absorption, we compute the correlation of its input singular vector $\bf{v_1}$ with the one without absorption $\bf{v_0}$:
\begin{eqnarray}
C_T = |( \bf{v_0,v_1} )| , \label{eqn_corr}
\end{eqnarray}
where $(\bf{v_0,v_1}) = \bf{v_0}^{\dagger}\bf{v_1} $ is the inner product of the normalized singular vectors $\bf{v_1} $ and $\bf{v_0}$. 
Figure \ref{Figure_3} plots the channel's correlation $C_T$, averaged over 40 random realizations, as a function of $L/\xi_a$. 
Its value stays close to 1 when system length is smaller than the diffusive absorption length $L < \xi_a$, and it drops abruptly as $L > \xi_a$. 
In the same figure, we also plot the ensemble-averaged highest transmission eigenvalue $\langle \tau_1 \rangle$ versus $L/\xi_a$.
Due to the small number of input channels ($N = 20$) in the waveguide, the highest transmission eigenvalue $\langle \tau_1 \rangle$ does not reach 1 even without absorption ($L/\xi_a = 0$). 
As the absorption $L/\xi_a$ increases, the highest transmission eigenvalue $\langle \tau_1 \rangle$ decreases much earlier than the channel's correlation $C_T$. 
For example, at $L/\xi_a = 2.2$, the highest transmission eigenvalue $\langle \tau_1 \rangle$ is already reduced by a factor of 4, while the correlation $C_T$ remains more than 0.9.
Thus in the weak absorption regime, the maximal transmission eigenvalue decreases while the eigenvector remains almost the same. 
This means that interference remains strong, and absorption merely reduces the amount of energy reaching the output end, but does not change the interference pattern. 
However, in the strong absorption regime, the number of significant scattering paths is greatly reduced, and the interference effects are weakened. 
Consequently, the maximal transmission channel starts to change dramatically and becomes ``ballistic-like'' as we have seen in Fig. \ref{Figure_2}. \\

\subsection{\label{sec:sublevel2} Correlation of the maximal transmission channel with quasi-normal modes}  
\begin{figure}[htbp]
\centering
\includegraphics[scale=0.29]{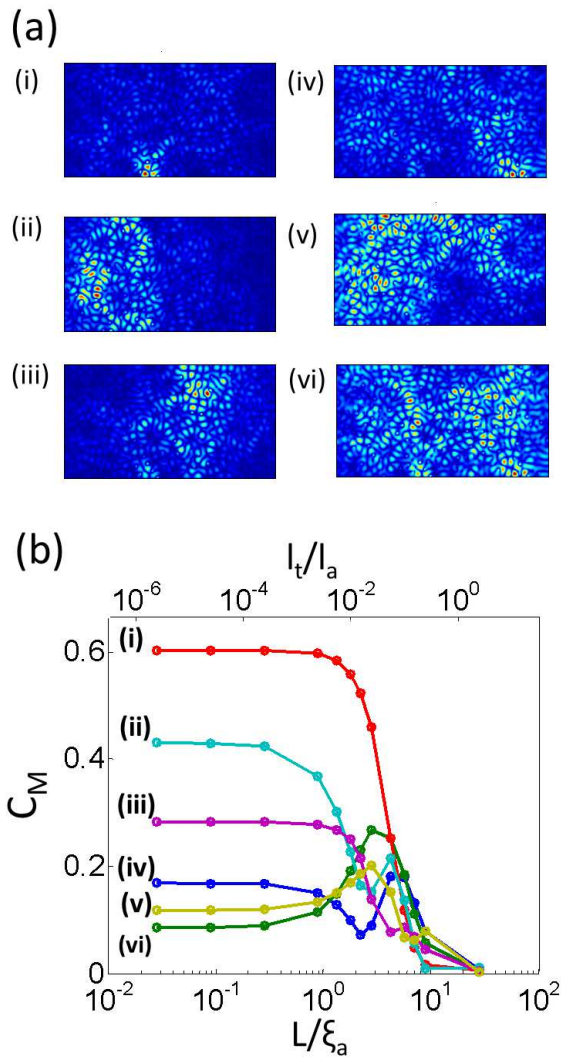}% Here is how to import EPS art
\caption{(Color online) Quasi-normal modes contributing to the maximal transmission channel. (a) Spatial distribution of electric field amplitude ($|E_z|$) for the six quasi-normal modes with the highest degree of correlation with the maximal transmission channel in Fig. \ref{Figure_2}.  (b) Correlation of the six quasimodes, labeled (i) - (vi) in (a), with the maximal transmission channel $C_M$ as a function of $L/\xi_a$. Modes that contribute the most when absorption is weak are spatially confined in the random structure and maximal transmission is facilitated through resonances hopping. }
\label{Figure_4}
\end{figure}
\indent To understand how the maximal transmission channel is formed and how it is modified by absorption, we investigate the related quasi-normal modes. 
Unlike the input or output singular vectors of the transmission matrix, the quasi-normal modes of an open random system are not orthogonal \cite{Leung_JPA97a, Leung_JPA97b, Ching_RMP98}, thus it is difficult to decompose the 2D field pattern of a transmission eigenchannel by the quasi-normal modes. 
Alternatively, we compute the degree of correlation between each quasimode and the eigenchannel to identify the modes that contribute significantly to the transmission channel. 
We used the commercial program Comsol to compute the complex frequency ($\omega = \omega_r + i \omega_i$) and field pattern of each quasi-normal mode in the disordered waveguide. 
The imaginary part $\omega_i$ of the complex frequency gives the decay rate or spectral width of the resonance, and the ratio of $\omega_r$ to $\omega_i$ is proportional to the quality factor. 
The contribution of a mode to the maximal transmission channel is reflected in the correlation of their spatial field profiles, $C_M = |\int E_q^*(x,y) E_c(x,y) dx dy|$, where $E_q(x,y)$ and $E_c(x,y)$ represent the normalized spatial distribution of the electric field $E_z$ for the mode and the channel, respectively. 
For the maximal transmission channel in Fig. \ref{Figure_2}, we identify six quasimodes with the highest degrees of correlation. 
Their frequencies are close to the frequency at which the transmission channels are computed, their field patterns are presented in Fig. \ref{Figure_4}(a). 
The first three modes, labeled (i) - (iii), have the dominant contributions to the maximal transmission channel at zero absorption. 
Mode (i) is a tightly confined mode, which is visible in the field profile of the eigenchannel [Fig. \ref{Figure_2}(a)]. 
Modes (ii) and (iii) are more extended, but they are not spread over the entire system, instead mode (ii) concentrates in the left-half of the disordered waveguide, and mode (iii) in the right-half. 
Their field patterns can be recognized in that of the eigenchannel. 
In contrast, modes (iv)-(vi) are spectrally further away from the probed frequency and are spatially extended over entire disordered waveguide. 
Since these three modes (i)-(iii) have little overlap in space, optimum transport of energy is facilitated by hopping through them. 
Hence, the maximal transmission channel can be regarded as a necklace of resonances strung from one side of the system to the other. 
It is similar to the necklace state that dominates the transport in the localization regime \cite{pendry_AP94}, except that it is not a single state (quasi-normal mode) but an eigenchannel of the transmission matrix.  
\\

\indent Figure \ref{Figure_4}(b) shows how the correlation between the maximal transmission channel and the quasi-normal modes change with absorption. 
In the regime of weak absorption ($L < \xi_a$), the correlation with each mode remains nearly constant, thus the field pattern of the transmission channel hardly changes. 
Note that the absorption is uniform and does not modify the spatial profile of individual quasimodes. 
When the absorption is strong ($L > \xi_a$), the correlations with modes (i)-(iii) decreases while the correlations with modes (v)-(vi) increases. 
These modes, unlike modes (i)-(iii), are spread more or less over the entire system. 
The resemblance in the field pattern between the maximal transmission channel at $L/\xi_a = 4$ in Fig. \ref{Figure_2}(a) and that of mode (v) in Fig. \ref{Figure_4}(a) indicates the mechanism of maximal energy transfer is changed from hopping through a series of resonances [modes (i)-(iii)] to delivering via an extended mode [mode (v)]. 
The reason for this change is that modes (i)-(iii) are more confined than (v), they have longer lifetime and are more attenuated by absorption.
Hence, in the presence of strong absorption, the maximal transport of energy is carried through extended modes with a shorter lifetime. 
With a further increase of absorption to $L/\xi_a = 9$, the channel does not resemble any particular quasimode. 
Light propagation through the random system is then dominated by ballistic-like transport, and interference no longer plays an important role in maximizing the transmittance. \\
                              
\subsection{\label{sec:sublevel3} Scaling of spectral width of maximal transmission channel with absorption}
\begin{figure}[!htbp]
\centering
\includegraphics[scale=0.19]{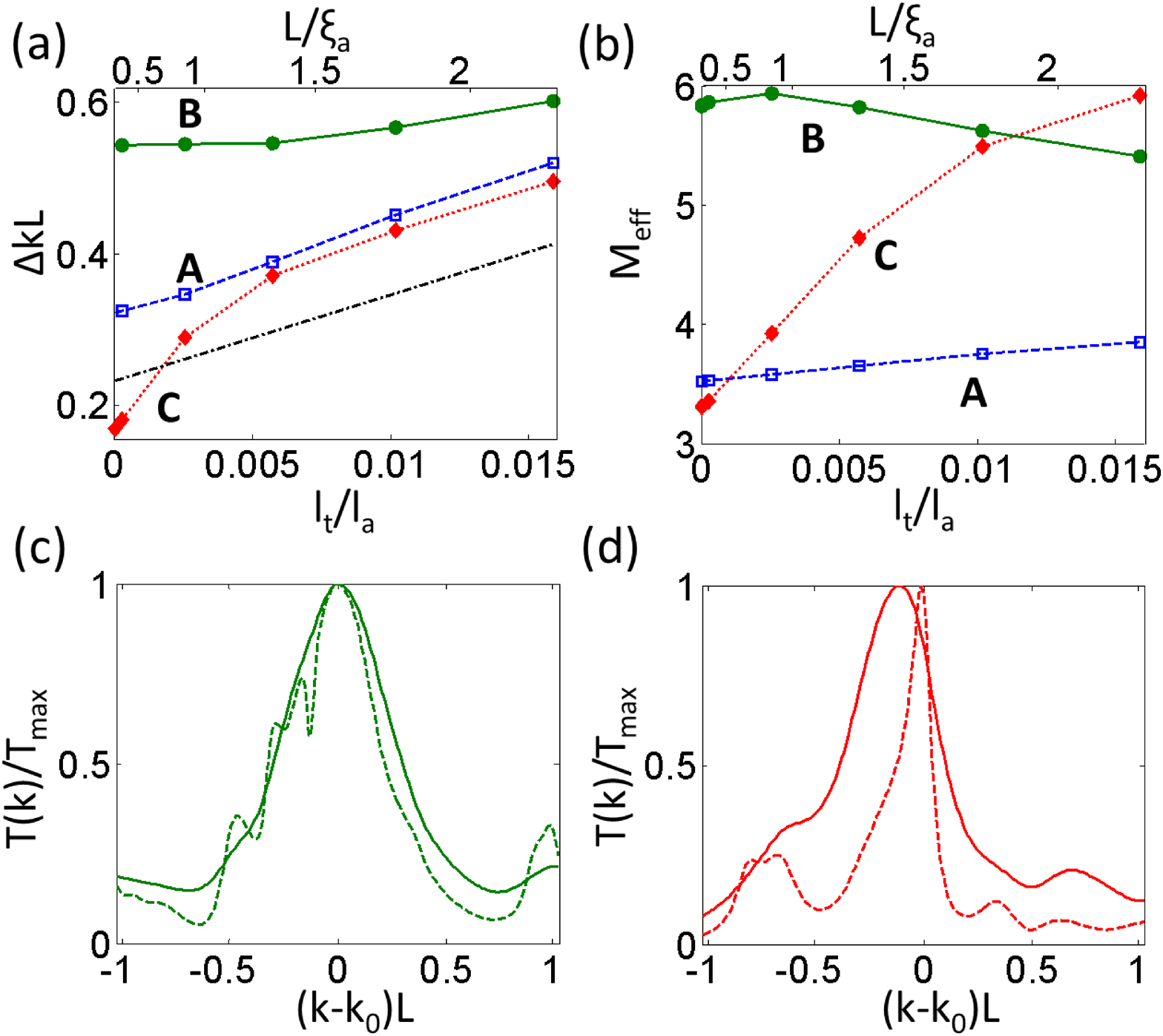}
\caption{(Color online) Dependence of the spectral width of the maximal transmission channel on absorption. (a) Spectral width of the maximal transmission channel $\Delta k$  with absorption (bottom axis $l_t/l_a$, top axis $L/\xi_a$) for three waveguide realizations A, B, C. The increase is linear for realization A (open squares connected by blue dashed line), sub-linear for B (filled circles connected by green solid line), and super-linear for C (filled diamonds with red dotted line). Black dash-dotted line represents the average linewidth for the quasimodes. (b) Mode participation number $M_{\rm eff}$ vs. absorption, it increases slightly for A, decreases for B and increases dramatically for C. (c,d) Total transmission for the input vector that gives the maximal transmission at $k_0$ as a function of frequency detuning of input light $k-k_0$ at $l_t/l_a = 0$ (dashed line) and 0.016 (solid line) for realizations B and C.}
\label{Figure_5}
\end{figure} 
 
\indent The profile of the maximal transmission channel changes with frequency. Its spectral width gives the frequency interval over which a fixed input wavefront, optimized at a single frequency, still leads to strongly enhanced transmission. 
A previous study on light focusing through lossless turbid media shows that the frequency bandwidth of a wavefront optimized for a single focus is equal to the width of speckle correlation function $D/L^2$, where $D$ is the diffusion coefficient \cite{beijnum_OL}. 
In this section, we investigate how absorption modifies the frequency bandwidth of the maximal transmission channel. \\

\indent To compute the bandwidth, we first input monochromatic light at frequency $k_0$ with the wavefront corresponding to the input singular vector of the maximal transmission channel, and calculate the total transmission $T(k_0)$ at the output end. 
Then, we scan the input light frequency $k$ while keeping the same wavefront and  calculate the new transmission value $T(k)$. 
As the frequency $k$ is detuned from $k_0$, the total transmission $T$ decreases. 
The bandwidth of the transmission channel is defined by the full-width-at-half-maximum (FWHM) as $\Delta k = k_2 - k_1$, where $T(k_1) = T(k_2) = T(k_0)/2$, $k_1 < k_0 < k_2$. 
With the introduction of absorption, $\Delta k$ increases. 
\\

\indent We observe varying scaling of $\Delta k$ with $l_t/l_a$ for various disorder configurations. 
Figure \ref{Figure_5}(a) shows three types of behavior where the bandwidth increases linearly for realization A (blue dashed line),  sub-linearly for realization B (green solid line), and super-linearly for realization C (red dotted line) with $l_t / l_a$. 
In contrast, all the quasi-normal modes exhibit the same linear increase of their spectral width with absorption.
The average spectral width of quasimodes is shown by the black dash-dotted line in Fig. \ref{Figure_5}(a). \\
%The average spectral width of quasimodes multiplied by the degree of spectral mode overlap or Thouless number\cite{Thouless} is shown by the black dash-dotted line in Fig. \ref{Figure_5}(a). 
%Thouless number is 1.67 in our system.

\indent To interpret these results, we again consider the quasi-normal modes underlying the transmission eigenchannel. 
We calculate the mode participation number defined as $M_{\rm eff} \equiv (\sum C_M)^2/ (\sum C_M^2)$. 
The summation includes modes within a fixed frequency range of $|k-k_0|L < 0.62$. 
Modes with a frequency beyond this range have a negligible contribution as they are spectrally located far outside the bandwidth of the maximal transmission channel. 
Figure \ref{Figure_5} (b) plots the mode participation number versus absorption for the three realizations A, B, C.
For A (linear scaling of $\Delta k$ with $l_t/l_a$), $M_{\rm eff}$ increases slightly with $l_t/l_a$, whereas for B (sublinear scaling of $\Delta k$ with $l_t/l_a$), $M_{\rm eff}$ decreases. In contrast, $C$, which features a superlinear scaling of $\Delta k$ with $l_t/l_a$, exhibits a rapid increase in $M_{\rm eff}$ as absorption increases.  
These results indicate that the bandwidth of the maximal transmission channel is related to the number of quasimodes contributing to the transmission. 
Figure \ref{Figure_5} (c) and (d) show the total transmission for the input vector that gives the maximal transmission at $k_0$ versus the frequency detuning of input light $k-k_0$ at $l_t/l_a = 0, 0.016$ for realizations B and C, respectively. 
We note that without absorption, the transmission of B that possesses a higher mode participation number, has a broader bandwidth than C. 
This is explained by the fact that more modes at different frequencies contribute to the total transmission.
The dramatic increase in the mode participation number for C adds to the absorption-induced broadening, leading to a super-linear increase of the channel bandwidth. 
Conversely, for B the broadening due to the increase of absorption is partially compensated by the reduction in the number of modes participating in the transmission, resulting in a sub-linear behavior.\\

\indent Even though there appears to be different scaling behavior from one disorder realization to another, the ensemble-averaged bandwidth of the maximal transmission channel increases linearly with absorption. 
%[following the black dash-dotted line in Fig. \ref{Figure_5}(a)]. 
This result echoes the finding reported in Ref. \cite{chong_PRL11} where the linewidth of perfect absorption channel exhibits a linear scaling with absorption. 

\section{\label{sec:level4} Minimal reflection channel}
\begin{figure}[htbp]
\centering
\includegraphics[scale=0.285]{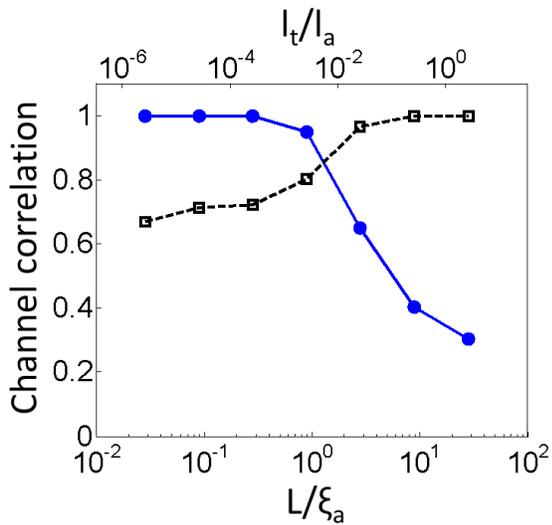}
\caption{(Color online) Absorption-induced change of minimal reflection channel. Filled circles connected by solid line represent the correlation between the minimal reflection channel and the maximal transmission channel. Open squares connected by dashed line represent the correlation between the minimal reflection channel and the maximal absorption channel. The minimal reflection channel is the same as the maximal transmission channel when absorption is weak $L < \xi_a$, but it approaches the maximal absorption channel as absorption is strong $L > \xi_a$. }
\label{Figure_6}
\end{figure}

In lossless random media, the maximal transmission channel is equivalent to the minimal reflection channel; the only way of reducing reflection is to enhance transmission. 
In an absorbing medium, this is no longer true: reflection may be reduced by enhancing absorption instead of transmission. 
In this section, we study the relations between the maximal transmission channel, the minimal reflection channel and the maximal absorption channel in absorbing random media. \\

\indent The reflection eigenchannels are obtained by singular value decomposition of the field reflection matrix $r$. 
The input singular vector corresponding to the lowest singular value gives the minimal reflectance. 
We compute the correlation between the input singular vector for maximal transmission and that for minimal reflection using the same definition as in Eq.(\ref{eqn_corr}) and present the result in Fig. \ref{Figure_6} (blue solid line). 
The correlation is almost one in the weak absorption regime $(L < \xi_a)$, but drops quickly once in the strong absorption regime $(L > \xi_a)$.
When the absorption is weak, the lowest reflection is still achieved by maximizing transmission. 
With strong absorption, light that is not reflected can be either transmitted or absorbed. 
Hence, reflection is reduced by enhancing both transmission and absorption. \\

\begin{figure}[!htbp]
\centering
\includegraphics[scale=0.24]{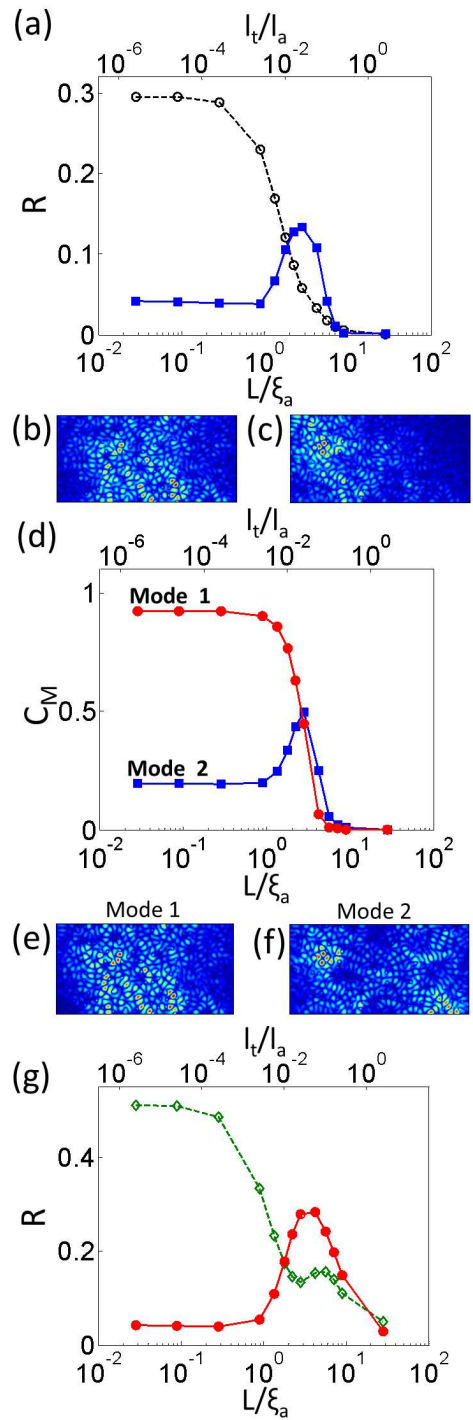}
\caption{(Color online) Example of increasing minimal reflectance with absorption. (a) Open circles connected by dashed line represent the ensemble-averaged lowest reflection eigenvalue as a function of absorption. Filled squares connected by solid line represent the lowest reflection eigenvalue of a selected realization that increases with absorption. (b-g) are for the selected realization. (b, c) Spatial distribution of the electric field amplitude $|E_z(x,y)|$ inside the random waveguide at $L/\xi_a = 0$ and $3$, respectively. The lowest reflection eigenvalue reaches a local maximum at $L/\xi = 3$. (d) Correlation of the minimal reflection channel with quasimodes 1 and 2. (e, f) Electric field patterns of mode 1 and 2. (g) Filled circles connected by solid line (open diamonds connected by dashed line) represent the reflectance of a fixed input wavefront $E_1$($E_2$) that corresponds to the input singular vector of the minimal reflection channel at $L/\xi_a = 0$ ($L/\xi_a = 3$). 
}
\label{Figure_7}
\end{figure}

\indent To find the maximal absorption channel, we introduce the matrix $h$ that links light incident from one end of the waveguide to the transmitted and reflected fields:
\begin{eqnarray}
h = \left( \begin{array}{ccc}
r \\
t \end{array} \right) 
\end{eqnarray}
where $r$ and $t$ are the field reflection and transmission matrices. 
An eigenvalue of $h^{\dagger}h$ represents the sum of the reflectance and the transmittance of its associated input singular vector.
The singular vector of $h$ with the smallest singular value corresponds to the maximal absorption channel. 
The correlation coefficient between the minimal reflection and maximal absorption channels is plotted as a function of the absorption in Fig. \ref{Figure_6} (black dashed line).
As the correlation between the minimal reflection channel and the maximal transmission channel decreases, the correlation between the minimal reflection channel and the maximal absorption channel increases. 
Eventually, in the strong absorption regime, the minimal reflection channel becomes identical to the maximal absorption channel, indicating that the minimal reflection is achieved by maximizing absorption instead of transmission.  \\

\indent Intuitively one expects all reflection eigenvalues to decrease with increasing absorption. 
Indeed, we show in Fig. \ref{Figure_7}(a), the ensemble-averaged minimal reflection eigenvalue decreases with absorption. 
Strikingly, in a significant number of realizations we have observed the opposite behavior, a counter-intuitive increase of reflection caused by absorption. 
In Fig. \ref{Figure_7}(a) we show the minimal reflectance of a selected realization. 
The minimal reflectance (eigenvalue of $r^{\dagger}r$) first decreases slightly as $L/\xi_a$ increases from 0 to 1, then rises rapidly by a factor of 3 before dropping again at larger absorption. 
In Fig. \ref{Figure_7}(b) and (c) we show the electric field patterns inside the random waveguide corresponding to the minimal reflectance without absorption and at $L/ \xi_a = 3$ where the lowest reflection eigenvalule reaches a local maximum. 
In the absence of absorption, light penetrates deep into the random medium with the field maxima close to the center of the sample. 
However, at $L/\xi_a = 3$, the penetration depth is greatly reduced and the field maxima shift to the input end of the random system. 
The field pattern close to the input surface of the same is also strongly modified. \\

\indent To understand this counter-intuitive behavior, we investigate the quasi-modes that contribute to the minimal reflection channel. 
Fig. \ref{Figure_7}(d) plots the degree of correlation between the minimal reflection channel and two quasi-modes (labeled 1, 2) that have the highest contributions, and in Fig. \ref{Figure_7}(e) and (f) show their field patterns. 
Without absorption, the least reflection channel is dominated by mode 1 that is located near the center of the sample. 
Destructive interference of various scattering paths of light in the disordered waveguide minimizes the reflectance. 
This explains the low minimal reflectance value of the selected realization compared to the ensemble-averaged one. 
With the introduction of absorption to the system, relative amplitudes of these paths are changed, the longer paths are attenuated more than the shorter paths, thus the destructive interference is weakened, leading to an increase of the reflectance as shown in Fig. \ref{Figure_7}(a). 
When the minimal reflectance has a maximum at $L/\xi_a = 3$, the field pattern of the minimal reflection channel shown in Fig. \ref{Figure_7}(c), resembles that of mode 2 in the left half of the waveguide. 
At this strong absorption level, the contribution of mode 1 to the minimal reflection channel is negligible, whereas mode 2 becomes dominant, which is a mode with a larger overlap with the input light into the disordered waveguide. 
Therefore, the interference in the bulk of the sample - that is the cause of open transmission channels - is suppressed, which leads to an increased reflectance. \\

\indent Let us now consider the input wavefront $E_1$, which corresponds to the minimal reflection channel (or maximal transmission channel) without absorption. 
$E_1$ couples most of the input energy into mode 1. 
Figure \ref{Figure_7}(g) shows the reflectance associated with input wavefront $E_1$ as a function of absorption $L/\xi_a$. 
When absorption becomes significant ($L > \xi_a$), the reflectance increases rapidly, up to $5$ times its value at $L/\xi_a =0$. 
Another input wavefront $E_2$, corresponding to the minimal reflection channel at $L/\xi_a = 3$, couples most of the energy into mode 2. 
In Fig. \ref{Figure_7}(g), we show the evolution of the reflection value associated to this wavefront with absorption. 
As expected, the reflectance decreases with increasing absorption. 
Once the reflectance with input wavefront $E_2$ becomes lower than that associated with $E_1$, mode 2 becomes dominant in the minimal reflection channel. 
This illustrates that with increasing absorption, the mechanism that produces the minimal reflection channel changes from transmission to bulk absorption. 
Moreover, it shows the significant role played by interference of scattered waves up to the point where the absorption becomes dominant.  \\

\section{\label{sec:level5} Conclusion}
We have performed a detailed numerical study to understand how absorption modifies transport in a 2D disordered waveguide, with emphasis on the maximum transmission channel. 
The maximal transmission channel is relatively robust against absorption compared to the transmittance. 
Its input wavefront remains nearly unchanged up to $L \approx \xi_a$, but changes rapidly beyond that point. 
In the maximal transmission channel, light propagates through the random structure along winding paths when absorption is weak ($L < \xi_a$), and takes more straight routes once absorption is significant ($L > \xi_a$). 
We investigate the correlations between the quasi-normal modes and the maximal transmission channel to illustrate the mechanism of enhanced transmission in both weak and strong absorption regimes. 
Maximal transmission is facilitated by hopping through localized modes when absorption is weak, and is dominated by more extended modes when absorption is strong.
We observe distinct scaling behavior for the spectral width of maximal transmission channel in different random configurations. Such differences result from the absorption-induced change in the number of quasimodes that participate in the maximal transmission channel.  
The channel spectral width increases linearly with absorption, if the mode participation number $M_{\rm eff}$ remains almost constant; the width increases sublinearly if $M_{\rm eff}$ decreases, and superlinearly if $M_{\rm eff}$ increases. 
In the absence of absorption, minimal reflection corresponds to maximal transmission, but this correspondence no longer holds in the regime of strong absorption ($L > \xi_a$), where minimal reflection corresponds to enhanced absorption. 
In some instances, we have observed the surprising feature that the minimal reflection eigenvalue increases with absorption, which can be explained by the reduction of destructive interference. 
The numerical study presented here provides a physical understanding of the effects of absorption on transmission and reflection eigenchannels at the relevant mesoscopic scale, which will hopefully serve in the interpretation of experimental work, and in the design of practical applications. \\

\section{Acknowledgment}
We want to thank Douglas Stone, Arthur Goetschy, Alexey Yamilov for stimulating discussions and Ad Lagendijk for encouragements. The numerical study is supported in part by the facilities and staff of the Yale University Faculty of Arts and Sciences High Performance Computing Center. This work is funded by the US National Science Foundation under the Grant Nos. ECCS-1068642 and DMR-1205307, and by ERC(279248), FOM(``Stirring of Light''), NWO, STW.

\end{document}